# Heat Diffusion in High-$C_p$ Nb$_3$Sn Composite Superconducting Wires

**Emanuela Barzi [1,\*], Fabrizio Berritta [2], Daniele Turrioni [1], and Alexander V Zlobin [1]**

[1] Fermi National Accelerator Laboratory, Pine and Kirk Rds. - Batavia 60510, IL, US; barzi@fnal.gov, turrioni@fnal.gov, zlobin@fnal.gov

[2] Politecnico di Torino, Department of Electronics and Telecommunication, Corso Duca degli Abruzzi, 24, 10129 Torino; fabrizio.berritta@gmail.com

\* Correspondence: barzi@fnal.gov

**Abstract:** A major focus of Nb$_3$Sn accelerator magnets is on significantly reducing or eliminating their training. Demonstration of an approach to increase the $C_p$ of Nb$_3$Sn magnets using new materials and technologies is very important both for particle accelerators and light sources. It would improve thermal stability and lead to much shorter magnet training, with substantial savings in machines' commissioning costs. Both Hypertech and Bruker-OST have attempted to introduce high-$C_p$ elements in their wire design. This paper includes a description of these advanced wires, the finite element model of their heat diffusion properties as compared with the standard wires, and whenever available, a comparison between the minimum quench energy (*MQE*) calculated by the model and actual *MQE* measurements on wires.

**Keywords:** Superconducting Nb$_3$Sn; specific heat; ceramic powders; accelerator magnets; superconducting magnet training

## 1. Introduction

A major focus of Nb$_3$Sn high field accelerator magnets for High Energy Physics (HEP) is on significantly reducing or eliminating their training by understanding its underlying physics mechanisms [1]. Superconducting magnets quench when their temperature increases above the current sharing temperature of the composite superconductor over a large enough volume. The temperature increase $\Delta T$ is proportional to $\Delta Q/C_p$, where $\Delta Q$ is the dissipated heat, and $C_p$ is the volumetric heat capacity. Energy deposition that initiates quenches can emanate from a variety of sources (flux jumps, conductor motion, epoxy cracking, etc). Another source of magnet training are materials and material interfaces, such as insulation, impregnating material, and neighboring structural materials. All these sources contribute to a resulting "disturbance spectrum".

Long training is a feature of any Nb$_3$Sn magnet. This includes both high field accelerator magnets developed and tested in the U.S. and abroad [2, 3], and superconducting undulator magnets under development for installation in the Advanced Photon Source (APS) storage ring at Argonne National Lab [4, 5]. In short models of Nb$_3$Sn accelerator magnets, the first quench (i.e. transition from superconducting to normal phase) generally occurs at 60-70% of the short sample limits and more than 20 quenches are required to reach the magnet nominal field [6]. In many cases, the number of quenches is proportional to magnet length, and training duration is expected to further increase for full-scale magnets.

The consistent increase of critical currents density in state-of-the-art Nb$_3$Sn wires has made the problem even more challenging. Perturbations result in local thermal heating of the wire that could eventually make the wire phase change to the normal state. In an adiabatic model, the minimum quench energy (*MQE*) is simply defined as the integration of the conductor's specific heat over the temperature margin within the superconducting state. The idea to reduce the *MQE* by inserting high specific heat (high-$C_p$) elements in superconducting wires dates to 1960 [7]. In the mid-2000s, a considerable improvement of stability to pulsed disturbances was obtained for NbTi windings, when distributing large heat capacity substances on the conductor during winding [8, 9]. The windings were brushed with $CeCu_6$ and $HoCu_2$ in the form of powders with 50-70 μm grain size and volumetric content of 3 to 6%. The powders were mixed with epoxy resin and the "wet-winding" process was used. The tests were performed with both small NbTi coils made of wire and larger windings made of Rutherford-type cable. The minimum quench energies of the mixed coils were several times higher than for the clean ones. It was found that the efficiency of the wet-winding with the large heat capacity powders was greatest for temperature diffusion times much smaller than the disturbance pulse duration. $Nb_3Sn$ is an intermetallic type II superconductor with a critical temperature $T_{c0}$ of 18.3 K and upper critical field $B_{c20}$ up to 30 T. For comparison, NbTi has a $T_{c0}$ of 9.8 K and $B_{c20}$ of 14.5 T. Demonstration of an approach to increase the $C_p$ of $Nb_3Sn$ magnets using new materials and technologies is very important both for particle accelerators and light sources. It would improve thermal stability and lead to much shorter magnet training, with substantial savings in machines' commissioning costs.

In standard Internal Tin $Nb_3Sn$ composite round wires, each superconducting subelement is composed of a Sn rod surrounded by Nb, enclosed in a Cu sheath. In the Restack-Rod-Process (RRP®) by Bruker-OST, the Nb is itself composed of dozens of microfilaments [10]. In the Tin-in-Tube by Hypertech, the Nb is in the form of hexagonal tubes [11]. In both cases, the superconducting subelements are positioned in a normal low resistance matrix to ensure some stability with respect to flux jumps and protection in case of quenching. Once the hexagonal Nb-Sn-Cu subelements and Cu rods are stacked into a Cu can to compose a billet, the billet is then drawn down into a wire of a specific size. The composite Nb-Sn-Cu subelements become superconducting $Nb_3Sn$ and bronze replaces the Sn after high temperature heat treatment in inert gas. An example of a 150/169 wire cross section before heat treatment is shown in Fig. 1, left. The number 150 refers to the sum of superconducting elements in the wire, whereas 169 is the maximum value that could be achieved without the internal Cu hexagon. The main advantages of the Cu matrix are the high thermal conductivity and the high specific heat. The former enhances heat transfer away from the filaments, while the latter promotes the absorption of a large fraction of heat and decreases Joule heating as the superconductor loses its superconductor capability. Induced eddy currents by time-dependent fields are reduced by twisting the filaments; this solution also improves stability to flux jumps. To obtain the required current in a Rutherford cable, several strands are connected in parallel and twisted or transposed along the axial direction (Fig. 1, right). The strands in the Rutherford cable are not insulated from each other for letting the current redistribute in the case of localized defects or quenches [12].

Both Hypertech and Bruker-OST have attempted to introduce high-$C_p$ elements in their wire design [13]. This paper includes a description of these advanced wires, the finite element model of their heat diffusion properties as compared with the standard wires, and whenever available, a comparison between the *MQE* calculated by the model and actual *MQE* measurements on wires.

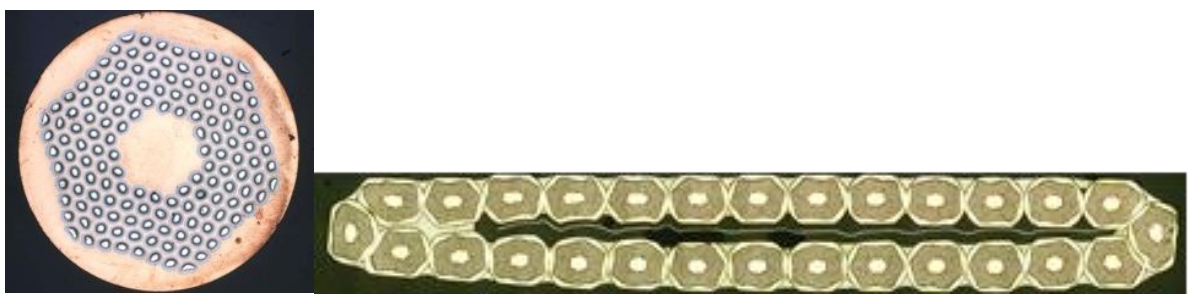

**Figure 1.** Left: Cross section before heat treatment of 150/169 $Nb_3Sn$ wire of the Restacked Rod Process® type [10]. Right: Cross section of 28-strand Rutherford cable [12].

## 2. Wire Sample Description

Examples of the two high-$C_p$ composite wires that were modeled for thermal diffusion are shown in Fig. 2. In both cases, some of the superconducting and Cu hexagons in the billet were replaced by Cu tubes containing a mixture of Cu and $Gd_2O_3$ ceramic powders. It is known that $Gd_2O_3$ has a high specific heat at low temperatures, e.g. the monoclinic structure shows a peak at 3.8 K [14]. Cu powder is added to the composite to enhance thermal diffusivity among the high-$C_p$ particles and to soften the high-$C_p$ element for drawing. The Hypertech wire had 12 high-$C_p$ subelements, 6 larger at the center, and 6 smaller at the outer corners. The Bruker-OST wire had 24 high-$C_p$ subelements all placed on the outermost row to provide a more efficient interception of the heat flux from the wire surface. Hypertech used a Cu powder of -325 mesh and nano $Gd_2O_3$ powder, with a Cu:$Gd_2O_3$ mass ratio of 1:2. Bruker-OST experimented first within a range of Cu:$Gd_2O_3$ mass ratios and Cu tube thicknesses and eventually selected a Cu:$Gd_2O_3$ mass ratio of 4:1. After restacking and drawing, the Hypertech wire was produced in short lengths down to 0.7 mm, whereas the Bruker-OST billet began to break extensively at a wire size of 4.1 mm. Since the *MQE* of the Hypertech wire had been previously tested [13, 15], finite element modeling of the actual wire was used to verify the model against the data. The Bruker-OST layout was instead used to study the thermal efficiency of the high-$C_p$ tubes when arranged in a variety of different geometrical configurations.

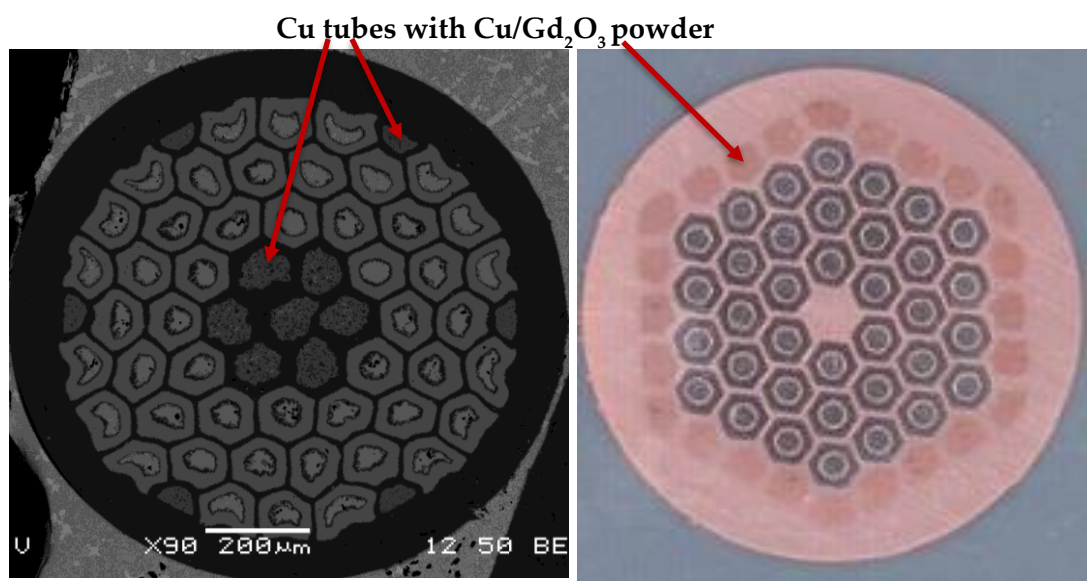

**Figure 2.** Left: Cross section before heat treatment of the high-$C_p$ 48/61 $Nb_3Sn$ wire of the Sn-in-Tube type by Hypertech [13]. Right: Cross section of the high-$C_p$ 36/61 $Nb_3Sn$ wire of the RRP type by Bruker-OST.

## 3. Experimental Setups and Methods

### 3.1. Sample Heat Treatment and Critical Current Measurement

To commission an upgraded *MQE* measurements system several $Nb_3Sn$ wire samples were used. In preparation to heat treatment in inert Argon, approximately 2 m long samples of $Nb_3Sn$ composite wire are wound on grooved cylindrical Ti-alloy (Ti-6Al-4V) barrels and held in place by two removable Ti-alloy end rings. The thermal cycles are performed in 3-zone controlled tube furnaces with a 12" long temperature homogeneity volume. Calibrated and ungrounded K-type thermocouples are used to monitor the accuracy and homogeneity of the reaction temperatures.

After heat treatment, the Ti-alloy end rings are removed from the Ti-alloy barrels and replaced by Cu rings. Voltage-current (*V-I*) characteristics are measured in boiling He at 4.2 K, in a transverse magnetic field. In standard wire critical current, or $I_c$, measurements, three pairs of voltage taps are used. Two pairs are placed along the center of the spiral sample 50 cm and 75 cm apart, and one pair at the Cu leads to be used for quench protection. The $I_c$ was determined from the *V-I* curve using the electrical field criterion of 0.1 µV/cm. Typical $I_c$ measurement uncertainties are within ±1% at 4.2 K and 12 T. The measured $I_c$ value is important when performing the *MQE* measurement at various normalized transport current ratios $I/I_c$.

*3.2 Minimum Quench Energy (MQE) Measurement and Reproducibility*

To measure the *MQE*, strain gauges are used as heaters. Strain gauges WK-09-125BT-350 from Micro-Measurements were glued to the samples using Stycast, with the gauge patterns (~4 mm in length and ~1.5 mm in width) centered on the wire sample and their long sides parallel to the wire axis. After curing of the Stycast, the instrumentation wires are soldered before sample and strain gauge get brushed with a thick layer (~1 mm) of Stycast. A 200 W KEPCO power supply provides the excitation voltage to the strain gauge. Using a LabView DAQ program, a µs-long pulse output is generated from the power supply and the voltage across the strain gauge is measured. With the $I_c$ of the sample first measured, a constant bias current below $I_c$ is applied to the sample and heat pulses are fired using the strain gauge. In this study, the duration of the heat pulses was fixed at 200 µs. A separate quench protection monitored the voltage across the sample and shut down the power supply if the quench threshold was reached. By gradually increasing the pulse power (in steps of 0.03 µJ when approaching the critical energy), the critical energy that induced a quench is defined as the *MQE* of the sample. The measurement error is given by the difference between the *MQE* achieved value and the energy provided to the heater at the previous step.

For reproducibility, the most important factor is good thermal contact between the heater and the sample. It takes some practice to develop a reliable procedure in attaching the strain gauge to the wire. Eventually, a very good reproducibility was achieved, as can be seen from Fig. 3, which shows the measured *MQE* for a few wires as a function of the normalized transport current $I/I_c$. Measurement errors as described above are shown in Figure. The 108/127 RRP wire in the plot is the standard wire used in the LHC Accelerator R&D Program (LARP), which is currently in its production phase as US HL-LHC Accelerator Upgrade Project (AUP). In addition, two samples of the same 60/61 RRP wire were flat rolled before reaction from 0.7 mm to 0.56 mm (i.e. 20% deformation) to allow for a better mounting of the strain gauge.

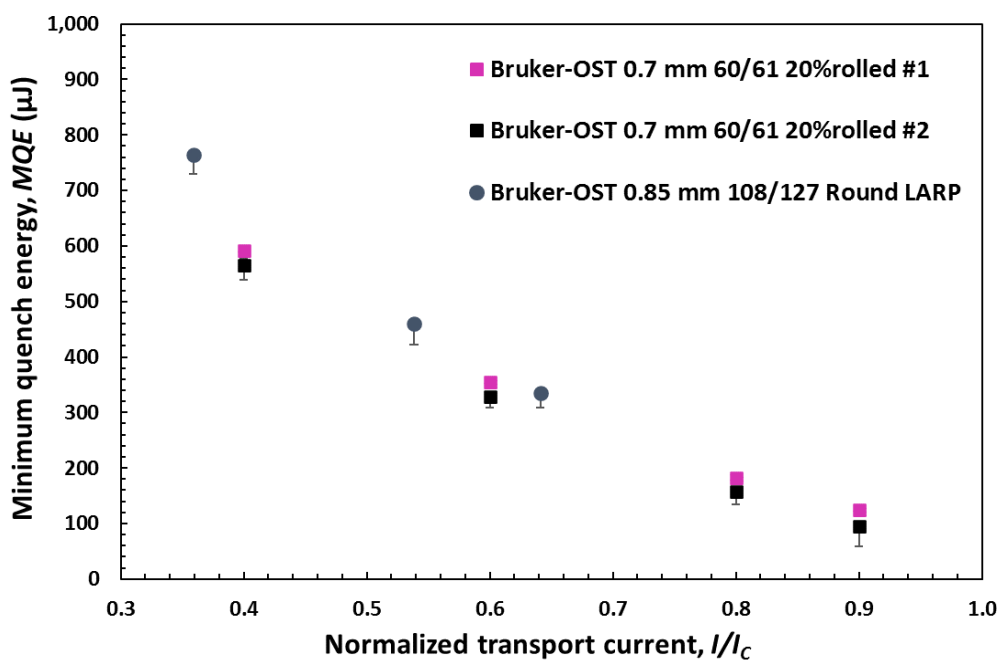

**Figure 3.** *MQE* as measured for a number of wires as function of the normalized transport current $I/I_c$.

## 4. Finite Element Model

*4.1 Quench Origin Mechanisms*

Critical current, temperature and magnetic field define a critical surface within which the superconducting state is sustained, i.e. non measurable resistivity and perfect diamagnetism. In general, a magnet operates below the critical surface, but as soon as the current is being ramped up one gets closer to the critical surface. Crossing it means that a small volume *V* switches to the normal resistive state. Power is therefore dissipated by Joule effect, leading to a temperature increase of the surrounding volume *dV*. If the temperature reached by *dV* approaches the critical temperature, then further power is dissipated, and the process keeps going by thermal diffusion. The normal zone propagates through the entire coil and the magnet quenches.

Two different quench mechanisms are defined: *conductor-limited* and *energy-deposited* quenches [16]. To distinguish between them, a conductor of known critical current $I_c(T,B)$ is assumed. The maximum magnetic field seen by the conductor is a function of the current itself $B = f(I)$, resulting in the following implicit equation for the maximum current $I_{max}$:

$$I = I_c(T_0, f(I)) \; at \; T_0 \; .$$

If the quench occurs at $I_{max}$, then it is due to the intrinsic properties of the conductor: it is a conductor limited quench. The other type of quench occurs at a current *I* lower than $I_{max}$ ($T_0$), once that the temperature raises to $T_0 + \Delta T$ in a volume *dV* of the coil, such that

$$I \geq I_c(T_0, f(I)).$$

These quenches are energy-deposited since they take place because of an energy deposit. Hence, in conductor-limited quenches the critical surface is crossed because of a current increase, whereas in the energy-deposited quenches the critical surface is crossed due to a local temperature increase. In the following, we will consider only the latter since they are the cause of magnet training. Furthermore, the magnetic field *B* will be assumed to be just the external field provided by the solenoidal experimental setup.

*4.2 Materials' Properties*

The Finite Element Model (FEM) of heat diffusion in $Nb_3Sn$ wires was made using ANSYS Mechanical APDL®. With material properties dependent on temperature, the ANSYS model is non-linear, and iterative algorithms are required to obtain a solution with the desired accuracy [17]. Standard and high-$C_p$ wires by Hypertech and Bruker-OST were modeled in 2-D as depicted in Fig. 4. The reference system is assumed to be centered at the wire center, with the *x* and *y* axes as shown in Figure. Half of the wire geometry was modeled because of symmetry with respect to the *y* direction. The normal wire is surrounded by the Stycast, which acts as a thermal insulator, whereas the $Nb_3Sn$ subelements are embedded in a Cu matrix. The wire is not at the center of the Stycast to better represent the experimental setup. Since the wires are modeled after undergoing the thermal cycle, the $Nb_3Sn$ hexagons contain bronze, rather than Sn.

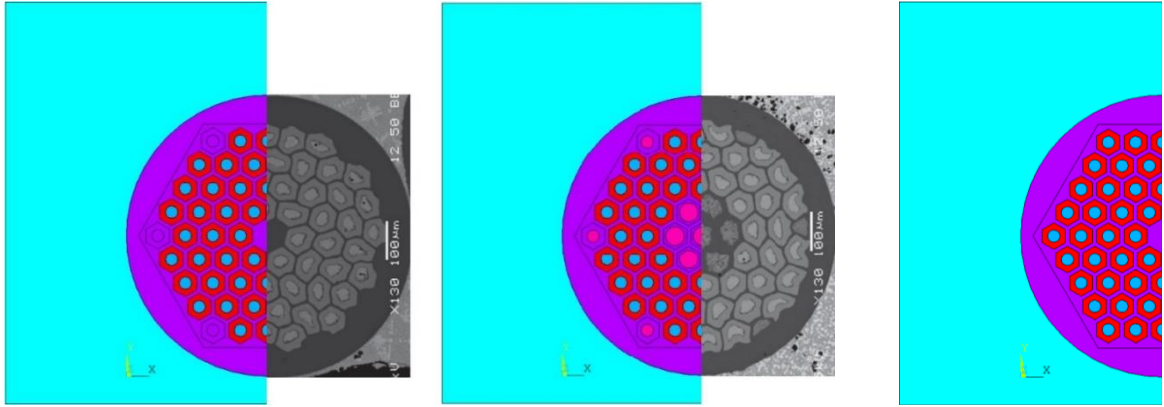

**Figure 4.** Left: Geometrical model used for standard 54/61 wire by Hypertech, compared with a cross section of the latter. Center: Geometrical model used for high-$C_p$ 48/61 wire by Hypertech, compared with a cross section of the latter. Right: Geometrical model used for standard 60/61 wire by Bruker-OST.

**Table 1.** Material properties: thermal conductivity $\kappa$ [W/(m K)], specific heat $C_p$ [J (kg K)] and density $d$.

| | $Nb_3Sn$ | | Cu | | Bronze (Sn wt%=5.46) | |
| --- | --- | --- | --- | --- | --- | --- |
| | [$d$ = 8400 kg m$^{-3}$] | | [$d$ = 8960 kg m$^{-3}$] | | [$d$ = 8850 kg m$^{-3}$] | |
| T | $\kappa$ [19] | $C_p$ [20] | $\kappa$ [20] | $C_p$ [20] | $\kappa$ [21] | $C_p$ [21] |
| 4 K | 174 | 0.41 | 158 | 0.091 | 1.9 | 0.129 |
| 6 K | 237 | 0.94 | 237 | 0.226 | 2.9 | 0.194 |
| 8 K | 308 | 1.85 | 315 | 0.470 | 3.9 | 0.387 |
| 10 K | 320 | 3.27 | 394 | 0.856 | 4.9 | 0.968 |
| | Stycast | | $Gd_2O_3$ | | | |
| | [$d$ = 2400 kg m$^{-3}$] | | [$d$ = 7410 kg m$^{-3}$] | | | |
| T | $\kappa$ [22] | $C_p$ [23] | $\kappa$ [24] | $C_p$ [18] | | |
| 4 K | 0.07 | 0.44 | 6.2 | 22 | | |
| 6 K | 0.11 | 1.70 | 6.2 | 27 | | |
| 8 K | 0.15 | 3.70 | 6.2 | 29 | | |
| 10 K | 0.19 | 6.20 | 6.2 | 29 | | |

The chosen material properties as a function of temperature are detailed in Table 1. The Nb$_3$Sn properties in the Table refer to the superconducting state. The effect of the applied magnetic field, *B*=12 T, is considered only for the Cu thermal conductivity and for the Gd$_2$O$_3$ specific heat, as shown in Fig. 5. All the Gd$_2$O$_3$ properties have been mixed with those of Cu to consider the powder mixture. The standard linear rule of mixture was used, with a Cu to Gd$_2$O$_3$ ratio of 1:2. It is interesting to notice in Fig. 5 that the peaking behavior of the specific heat is lost as magnetic fields are applied [18]. Nonetheless, the specific heat of Gd$_2$O$_3$ remains sufficiently high as experimental results show.

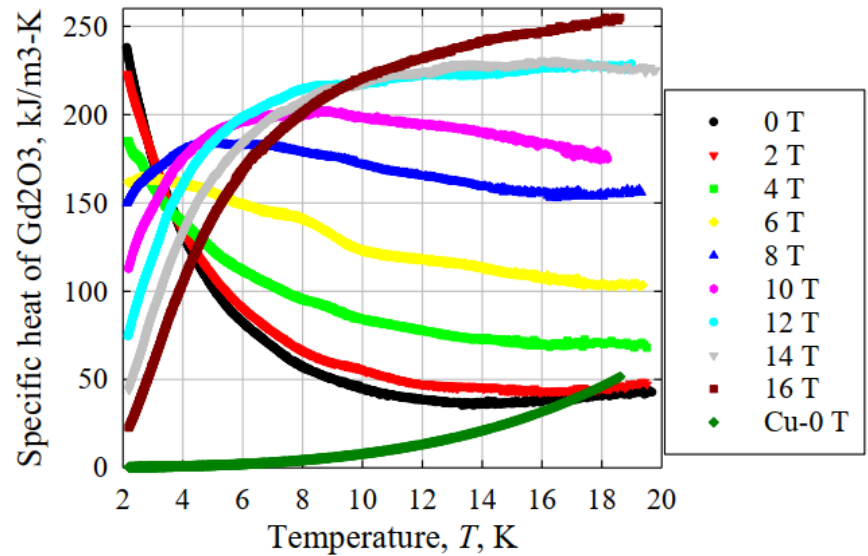

**Figure 5.** Measured specific heat of $Gd_2O_3$ [18].

*4.3 Wire Geometries*

As mentioned in Section 2, Hypertech and Bruker-OST standard and high-$C_p$ wires were modeled. The FEM model was first checked against measurements of the modeled Hypertech wires. Once the model's prediction capability was verified, it was applied to several different high-$C_p$ subelement layouts in the Bruker-OST wire.

4.3.1. Hypertech Wires

For the standard and high-$C_p$ wires by Hypertech in Fig. 4, left and center, the following geometrical parameters were used:

- Wire diameter = 0.7 mm
- Bronze rods radius = $Gd_2O_3$ corner subelement radius = 15 µm
- Maximum superconducting subelement width = 2×29/√3 µm
- Thickness between two sub-elements = 10 µm
- Stycast total width $W_{sty}$ and height $H_{sty}$ are 1.3 mm and 1.0 mm, respectively. Stycast center was displaced along *y* by $(H_{sty} − R_{wire})/8$
- $Gd_2O_3$ central subelement radius = 23 µm

4.3.2. Bruker-OST Wires

For the standard wire by Bruker-OST in Fig. 4, right, the same geometrical parameters as above were used. In this case, all the high-$C_p$ sub-elements have the same area as the bronze rods. In addition to the standard wire, six additional geometries were simulated:

- 6 high-$C_p$ sub-elements in the innermost row
- 6 high-$C_p$ sub-elements at the six outermost corners
- 24 high-$C_p$ sub-elements in the outermost row
- 24 high-$C_p$ sub-elements in the first two innermost rows and six outermost corners
- 18 high-$C_p$ sub-elements in the first two innermost rows
- 20 high-$C_p$ sub-elements randomly positioned

*4.4 Thermal Analysis*

The next step was the choice of the thermal load to simulate an *MQE* experiment. In [15] a flat strain gauge was applied on top of the wire and the energy was obtained by integrating over time the dissipated power due to Joule heating in the gauge. In the 2-D FEM model, a thermal flux impinging on a circumference arc was chosen, corresponding to an angle $\theta \in [90°, 120°]$. Since we are working with half of the wire geometry, this means that the heat flux was applied to $30° \times 2 / 360° = 1/6$ of the wire circumference. It will be shown that the choice of $\theta$ changes the results by small percentages. As a boundary condition, the initial temperature is set to be $T_0 = 4.2$ K because of the He bath, whereas at the boundaries $T_{env} = 4.2$ K is kept fixed during the transient. Along $x = 0$ no boundary conditions have been applied, resulting in an adiabatic setting for ANSYS. No heat flux occurs along that line because of the symmetry. The chosen element type was PLANE55, and the mesh was sufficiently refined until no changes occurred in the second decimal digit of the wire's maximum temperature after the heat pulse.

The heat flux pulse duration was set to 200 µs and its amplitude was evaluated by dividing the input energy by two, by the pulse duration and by the area. The factor of two derives from modeling half of the wire, whereas the area is just the arc length upon which the flux is applied multiplied by a unitary length, which is the thickness dimension used by ANSYS in 2-D models [17]. The goal of the model is to obtain the *MQE* as a function of the normalized transport current $I/I_c$. However, whereas in the experimental setup the current is the input and the *MQE* is then measured at that specific current, in this FEM model their roles are exchanged. In the FEM simulation no currents are available, just the maximum temperature of the wire is. The concept used was to consider the maximum temperature, at a given input energy, as the critical temperature associated with the required critical current. An empirical relationship for the $I_c(B,T)$ of Nb$_3$Sn was used [25] with the fitting parameters $I_{ref}(\varepsilon) = 489$ A, $T_{c0}(\varepsilon) = 16.96$ K and $B_{c20}(\varepsilon) = 27.89$ T, leaving the strain undetermined. With this method, the critical current $I_{c0} = I_c(12T, 4.2\ K)$ was obtained, as well as the current ratio $\tilde{I}_{Tc} = I_c(12T, T_c) / I_{c0}$ of the wire at the critical current $T_c$ for each heat flux pulse. For Nb$_3$Sn at 12 T, critical temperatures of 6.3 K and 4.4 K corresponded to critical current ratios of 0.2 and 0.8, respectively.

*4.5 Sensitivity Analysis*

To study the sensitivity of the model with respect to the chosen material properties, the FEM was tested several times, by changing the specific heat or thermal conductivity of each material separately. Each material property was varied by ±20%. The results are summarized in Tables 2 and 3. In the Tables, 'Dev.' stands for 'Deviation'. The minimum and maximum percentage change of $\tilde{I}_{Tc,\alpha i=\pm20\%}$, where $\alpha_i = \pm20\%$ represents a material property change at all temperatures, are indicated next to the associated input *MQE* values. It resulted that for each material property change, the minimum absolute $\tilde{I}_{Tc}$ deviation from the original simulation occurs at the lowest *MQE* input value, and the maximum absolute $\tilde{I}_{Tc}$ deviation at the maximum *MQE* input value. It is interesting to note that variations in the Nb$_3$Sn specific heat have the greatest impact on the standard wire final temperature, whereas for high-$C_p$ wires it is the Gd$_2$O$_3$ specific heat that has the greatest impact.

Similarly, the FEM was tested also when changing the angle $\theta$ representing the extension of the thermal load around the wire. As detailed in Table 4, the angle was changed by up to ±20°, and the $\tilde{I}_{Tc}$ changes by less than 5%. Smaller angles result in lower current ratio values. This behavior can be explained with a reduced heat transfer surface at small angles, resulting in lower heat transfer towards the Stycast and therefore in a slightly higher increase in the wire temperature.

**Table 2.** Materials' properties sensitivity in thermal simulation of standard wire.

| Thermal conductivity | $I/I_c$ Min. Dev. [%] | Min. $MQE$ [µJ] | $I/I_c$ Max. Dev. [%] | Max. $MQE$ [mJ] |
|---|---|---|---|---|
| Stycast: $\Delta\kappa = -20\%$ | -1.4 | $3.2 \cdot 10^2$ | -3.5 | $2.3 \cdot 10^3$ |
| Stycast: $\Delta\kappa = 20\%$ | 1.2 | $3.2 \cdot 10^2$ | 3.3 | $2.3 \cdot 10^3$ |
| Cu: $\Delta\kappa = -20\%$ | -0.2 | $3.2 \cdot 10^2$ | -0.6 | $2.3 \cdot 10^3$ |
| Cu: $\Delta\kappa = 20\%$ | 0.1 | $3.2 \cdot 10^2$ | 0.5 | $2.3 \cdot 10^3$ |
| Nb$_3$Sn: $\Delta\kappa = -20\%$ | -0.1 | $3.2 \cdot 10^2$ | -0.3 | $2.3 \cdot 10^3$ |
| Nb$_3$Sn: $\Delta\kappa = 20\%$ | 0.1 | $3.2 \cdot 10^2$ | 0.2 | $2.3 \cdot 10^3$ |
| Bronze: $\Delta\kappa = -20\%$ | <0.1 | $3.2 \cdot 10^2$ | <0.1 | $2.3 \cdot 10^3$ |
| Bronze: $\Delta\kappa = 20\%$ | <0.1 | $3.2 \cdot 10^2$ | <0.1 | $2.3 \cdot 10^3$ |
| **Specific heat** | $I/I_c$ Min. Dev. [%] | Min. $MQE$ [µJ] | $I/I_c$ Max. Dev. [%] | Max. $MQE$ [mJ] |
| Stycast: $\Delta C_p = -20\%$ | -1.1 | $3.2 \cdot 10^2$ | -3.0 | $2.3 \cdot 10^3$ |
| Stycast: $\Delta C_p = 20\%$ | 1.0 | $3.2 \cdot 10^2$ | 2.8 | $2.3 \cdot 10^3$ |
| Cu: $\Delta C_p = -20\%$ | -1.7 | $3.2 \cdot 10^2$ | -4.9 | $2.3 \cdot 10^3$ |
| Cu: $\Delta C_p = 20\%$ | 1.6 | $3.2 \cdot 10^2$ | 4.8 | $2.3 \cdot 10^3$ |
| Nb$_3$Sn: $\Delta C_p = -20\%$ | -3.7 | $3.2 \cdot 10^2$ | -9.7 | $2.3 \cdot 10^3$ |
| Nb$_3$Sn: $\Delta C_p = 20\%$ | 3.3 | $3.2 \cdot 10^2$ | 9.3 | $2.3 \cdot 10^3$ |
| Bronze: $\Delta C_p = -20\%$ | -0.3 | $3.2 \cdot 10^2$ | -0.8 | $2.3 \cdot 10^3$ |
| Bronze: $\Delta C_p = 20\%$ | 0.3 | $3.2 \cdot 10^2$ | 0.8 | $2.3 \cdot 10^3$ |

## 5. Results and Discussion

*5.1 Model*

The results of the Hypertech wires' simulation are presented in Fig. 6, where they are compared with experimental data [15]. It can be appreciated that the model is conservative with respect to the experiment. However, it predicts correctly the $MQE$ ratio between the high-$C_p$ wire and the standard one. This ratio is about 3 for both model and data. Therefore, the thermal model reproduces well the relative thermal efficiency of the high-$C_p$ wire when compared to the standard one. Nonetheless, the $MQE$ curves from the model are lower than the acquired data. A major difference between experiment and model is the way that the heat is applied to the sample by the strain gauge. In the experiment, heat conduction occurs in all directions, whereas in the simulation the heat flux was directed towards the wire itself. Therefore, the heat required in the simulation to reach the critical temperature is lower than in the experiment.

Table 3. Materials' properties sensitivity in thermal simulation of high-$C_p$ wire.

| Thermal conductivity | $I/I_c$ Min. Dev. [%] | Min. *MQE* [µJ] | $I/I_c$ Max. Dev. [%] | Max. *MQE* [mJ] |
|---|---|---|---|---|
| Stycast: $\Delta\kappa = -20\%$ | -0.5 | $8.6 \cdot 10^2$ | -2.0 | $5.2 \cdot 10^3$ |
| Stycast: $\Delta\kappa = 20\%$ | 0.5 | $8.6 \cdot 10^2$ | 1.9 | $5.2 \cdot 10^3$ |
| Cu: $\Delta\kappa = -20\%$ | -0.4 | $8.6 \cdot 10^2$ | -1.5 | $5.2 \cdot 10^3$ |
| Cu: $\Delta\kappa = 20\%$ | 0.3 | $8.6 \cdot 10^2$ | 1.1 | $5.2 \cdot 10^3$ |
| Nb$_3$Sn: $\Delta\kappa = -20\%$ | -0.2 | $8.6 \cdot 10^2$ | -0.6 | $5.2 \cdot 10^3$ |
| Nb$_3$Sn: $\Delta\kappa = 20\%$ | 0.1 | $8.6 \cdot 10^2$ | 0.5 | $5.2 \cdot 10^3$ |
| Bronze: $\Delta\kappa = -20\%$ | <0.1 | $8.6 \cdot 10^2$ | <0.1 | $2.3 \cdot 10^3$ |
| Bronze: $\Delta\kappa = 20\%$ | <0.1 | $8.6 \cdot 10^2$ | <0.1 | $5.2 \cdot 10^3$ |
| Gd$_2$O$_3$: $\Delta\kappa = -20\%$ | <0.1 | $8.6 \cdot 10^2$ | <0.1 | $5.2 \cdot 10^3$ |
| Gd$_2$O$_3$: $\Delta\kappa = 20\%$ | <0.1 | $8.6 \cdot 10^2$ | <0.1 | $5.2 \cdot 10^3$ |
| **Specific heat** | $I/I_c$ Min. Dev. [%] | Min. *MQE* [µJ] | $I/I_c$ Max. Dev. [%] | Max. *MQE* [mJ] |
| Stycast: $\Delta C_p = -20\%$ | -0.5 | $8.6 \cdot 10^2$ | -1.8 | $5.2 \cdot 10^3$ |
| Stycast: $\Delta C_p = 20\%$ | 0.4 | $8.6 \cdot 10^2$ | 1.7 | $5.2 \cdot 10^3$ |
| Cu: $\Delta C_p = -20\%$ | -0.7 | $8.6 \cdot 10^2$ | -3.1 | $5.2 \cdot 10^3$ |
| Cu: $\Delta C_p = 20\%$ | 0.7 | $8.6 \cdot 10^2$ | 3.1 | $5.2 \cdot 10^3$ |
| Nb$_3$Sn: $\Delta C_p = -20\%$ | -1.4 | $8.6 \cdot 10^2$ | -5.5 | $5.2 \cdot 10^3$ |
| Nb$_3$Sn: $\Delta C_p = 20\%$ | 1.3 | $8.6 \cdot 10^2$ | 5.3 | $5.2 \cdot 10^3$ |
| Bronze: $\Delta C_p = -20\%$ | -0.1 | $8.6 \cdot 10^2$ | -0.4 | $5.2 \cdot 10^3$ |
| Bronze: $\Delta C_p = 20\%$ | 0.1 | $8.6 \cdot 10^2$ | 0.4 | $5.2 \cdot 10^3$ |
| Gd$_2$O$_3$: $\Delta C_p = -20\%$ | -5.6 | $8.6 \cdot 10^2$ | -13 | $5.2 \cdot 10^3$ |
| Gd$_2$O$_3$: $\Delta C_p = 20\%$ | 4.8 | $8.6 \cdot 10^2$ | 14 | $5.2 \cdot 10^3$ |

Table 4. Load angle sensitivity for thermal simulation. The angle refers to the FEM model, i.e. $\theta$ is between 0° and 180°. The original angle in simulations is 30°.

| Standard wire | $I/I_c$ Min. Dev. [%] | Min. *MQE* [µJ] | $I/I_c$ Max. Dev. [%] | Max. *MQE* [mJ] |
|---|---|---|---|---|
| $\Delta\theta = -20°$ | -0.5 | $3.2 \cdot 10^2$ | -2.0 | $2.3 \cdot 10^3$ |
| $\Delta\theta = -10°$ | -0.1 | $3.2 \cdot 10^2$ | -0.6 | $2.3 \cdot 10^3$ |
| $\Delta\theta = 10°$ | 0.1 | $3.2 \cdot 10^2$ | 0.6 | $2.3 \cdot 10^3$ |
| $\Delta\theta = 20°$ | 0.3 | $3.2 \cdot 10^2$ | 1.1 | $2.3 \cdot 10^3$ |
| **High-$C_p$ wire** | $I/I_c$ Min. Dev. [%] | Min. *MQE* [µJ] | $I/I_c$ Max. Dev. [%] | Max. *MQE* [mJ] |
| $\Delta\theta = -20°$ | -1.3 | $8.6 \cdot 10^2$ | -4.6 | $5.2 \cdot 10^3$ |
| $\Delta\theta = -10°$ | -0.5 | $8.6 \cdot 10^2$ | -1.8 | $5.2 \cdot 10^3$ |
| $\Delta\theta = 10°$ | 0.4 | $8.6 \cdot 10^2$ | 1.4 | $5.2 \cdot 10^3$ |
| $\Delta\theta = 20°$ | 0.7 | $8.6 \cdot 10^2$ | 2.5 | $5.2 \cdot 10^3$ |

For the absolute predictive model, also the minimum propagation zone (*MPZ*) should be considered. Indeed, an energy deposited quench occurs only if the local increase in temperature grows beyond a given volume $V_0$ inside the wire. A 1-D treatment of the problem can be found in [26]. From a qualitative point of view, since the strain gauge was ~ 4 mm long while in the ANSYS model unitary thickness is assumed, the required heat in the laboratory would be higher considering the *MPZ* effect.

Once the model's relative prediction capability was verified, it was applied to a number of different high-$C_p$ subelement layouts in the Bruker-OST wire. The results of the *MQE* simulations for the various Bruker-OST wire geometries are presented in Fig. 7. It is important to outline that two pairs of curves having the same number of high-$C_p$ subelements overlap each other. The first overlapping is for a total number of 6 high-$C_p$ subelements, and the other occurs in the case of 24 total high-$C_p$ subelements. Hence, these FEM simulations have shown that the quantity of these high-$C_p$ components has a much greater impact on their desired thermal effect than their location. The curves having the high-$C_p$ components far from the wire center are slightly more efficient. Nonetheless, it appears that the mechanical feasibility of high-$C_p$ composite wires decreases the further away from the billet center these brittle tubes are inserted in the billet. Thus, the simulations have shown that it is possible to develop new structures without worrying about thermal efficiency if the number of high-$C_p$ subelements is sufficiently high.

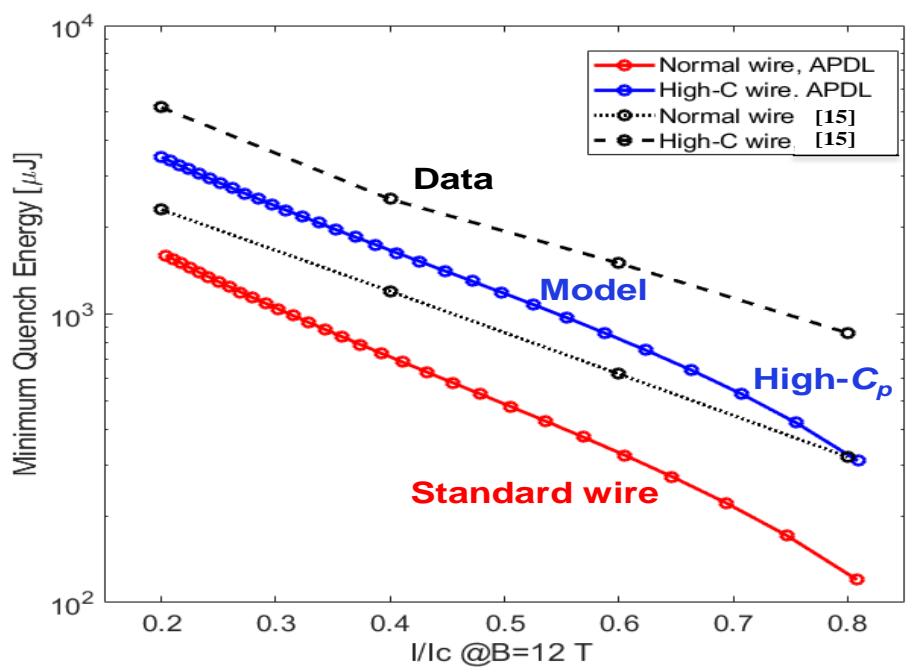

**Figure 6.** Comparison between *MQE* simulations and experimental data [15] for Hypertech wires.

*5.2 Experiment*

To verify the model against additional experimental data, a 0.85 mm Bruker-OST LARP standard wire was wrapped with a 10 mm wide and 64 μm thick $Cu/Gd_2O_3$ tape before heat treatment. *MQE* results, measured and simulated, are shown in Fig. 8 as a function of normalized transport current for the LARP-type wire bare and wrapped with the tapes. Results are comparable with the *MQE* achieved for the high-$C_p$ Hypertech wire.

## 6. Conclusions

Several 2-D FEM thermal models were developed to study the *MQE* of Nb₃Sn composite superconducting wires. The developed thermal model was verified with experimental data. It was found that the position of the high-specific heat subelements has a negligible effect on the thermal efficiency of the wire. The quantity of high-specific heat subelements, i.e. the fraction of cross-sectional area occupied by these subelements, is more important against thermal perturbation. This result offers greater freedom for the development of new high-$C_p$ composite wires and their mechanical feasibility.

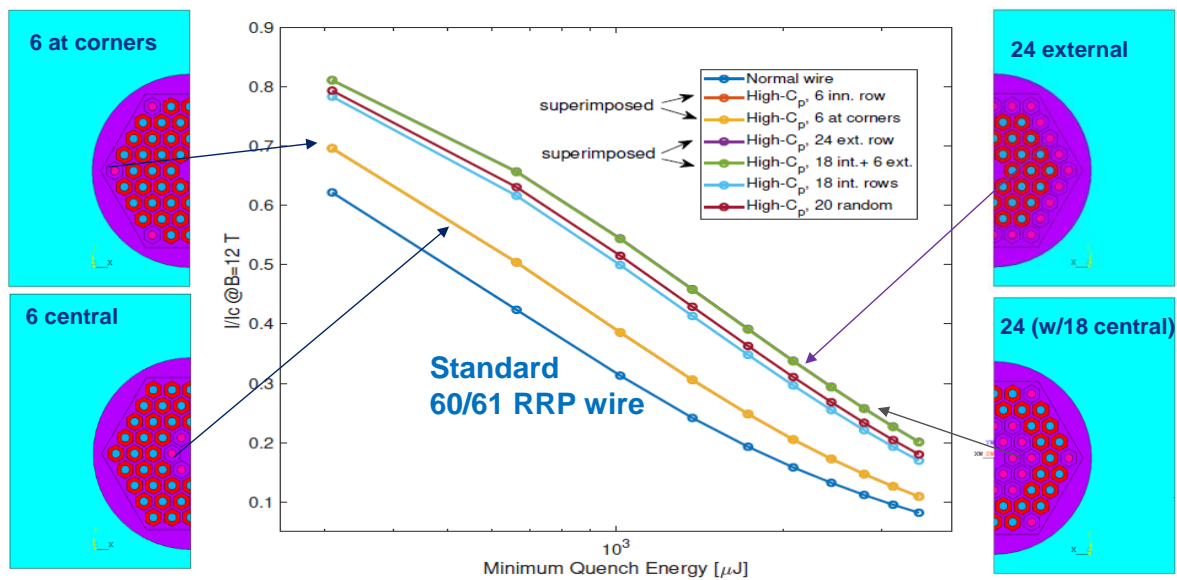

**Figure 7.** Thermal simulation of Bruker geometry with different high-$C_p$ subelements location and quantity.

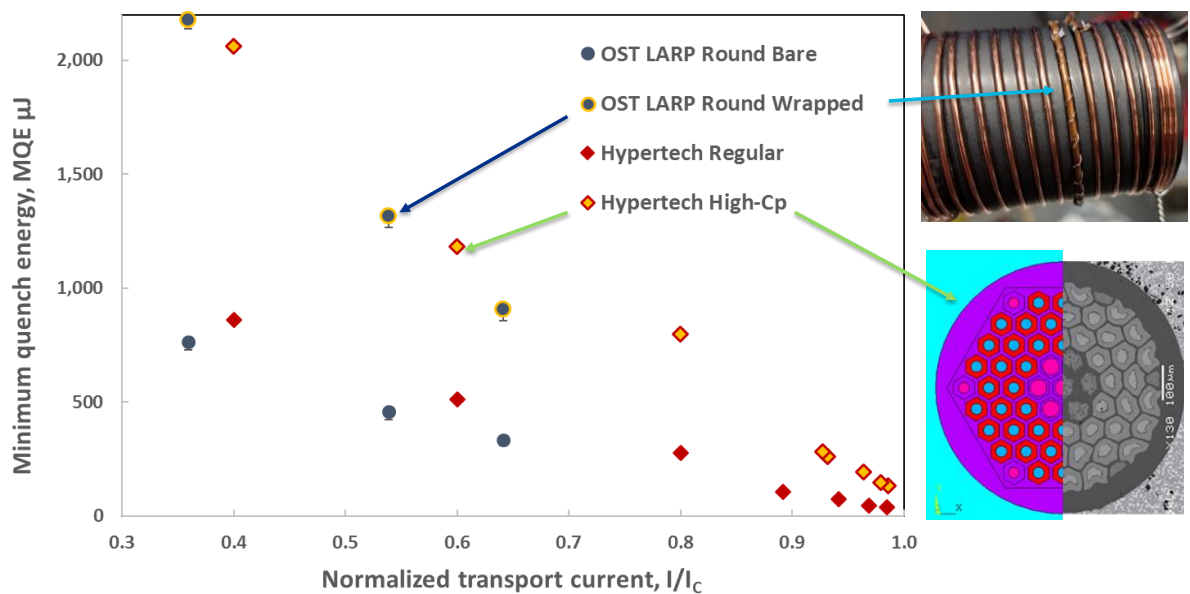

**Figure 8.** Comparison of *MQE* between high-$C_p$ material in tape form and with high-$C_p$ subelements within the strand geometry.

**Funding:** This work was supported in part by Fermi Research Alliance, LLC, under Contract no. DE-AC02-07CH11359 with the U.S. Department and by the US-MDP.

**Acknowledgments:** The authors thank the U.S. MDP collaboration and the Fermilab Summer School of the University of Pisa.

**References**

1. "The U.S. Magnet Development Program Plan," U.S. Department of Energy, Office of Science, June 2016.
2. "$Nb_3Sn$ Accelerator Magnets – designs, technologies, and performance," editors - D. Schoerling and A.V. Zlobin, ISBN 978-3-030-16117-0, Springer, 2019.
3. Ezio Todesco et al., "Progress on HL-LHC $Nb_3Sn$ Magnets," IEEE Trans. on Appl. Supercond., Vol. 28, Issue 4, June 2018, 4008809.
4. I. Kesgin et al., "Development of Short-Period $Nb_3Sn$ Superconducting Planar Undulator," I. Kesgin et al., IEEE Trans. on Appl. Supercond., vol. 29, No. 5, ASN 4100504, Aug. 2019.
5. I. Kesgin, M. Kasa, Q. Hasse, Yu. Ivanyushenkov, Y. Shiroyanagi, J. Fuerst, E. Gluskin (ANL) A.V. Zlobin, E. Barzi, D. Turrioni (FNAL), "Fabrication and Testing of 10-Pole Short-Period $Nb_3Sn$ Superconducting Undulator Magnets," IEEE Transactions on Applied Superconductivity, Volume 30, Issue 4, 2020, 10.1109/TASC.2020.2967686.
6. S. Stoynev, K. Riemer, A. V. Zlobin, G. Ambrosio, P. Ferracin, G.L. Sabbi and P. Wanderer, "Analysis of $Nb_3Sn$ Accelerator Magnet Training," IEEE Trans. on Appl. Supercond., Vol. 29, Issue 5, August 2019, 4001206.
7. R. Hancox, "Enthalpy stabilized superconducting magnets," IEEE Trans. Magn., vol. MAG-4, no. 3, pp. 486–488, Sep. 1968.
8. Alekseev P A et al., "Experimental evidence of considerable stability increase in superconducting windings with extremely high specific heat substances," 2004 Cryogenics 44 763-6.
9. Alekseev P A et al., "Influence of heat capacity substances doping on quench currents of fast ramped superconducting oval windings," 2004 Cryogenics 46 252-7.
10. M. B. Field et al., "Optimizing conductors for high field applications," IEEE Trans. Appl. Supercond, vol. 24, no. 3, p. 6001105, Jun. 2014.
11. X. Peng, E. Gregory, M. Tomsic et al., "Strain and Magnetization Properties of High Subelement Count Tube-type $Nb_3Sn$ Strands," IEEE Trans. Appl. Supercond, Vol. 21, No. 3, pp. 2559-2562, (2011).
12. E. Barzi and A. V. Zlobin, "Research and Development of $Nb_3Sn$ Wires and Cables for High-Field Accelerator Magnets," IEEE Transactions on Nuclear Science, vol. 63, no. 2, pp. 783-803, April 2016.
13. X Xu, P Li, A V Zlobin and X Peng, "Improvement of stability of $Nb_3Sn$ superconductors by introducing high specific heat substances," Supercond. Sci. Technol. 31 (2018) 03LT02.
14. R W Hill et al 1983 *J. Phys. C: Solid State Phys. 16 2871*.
15. X. Xu, A. V. Zlobin, X. Peng and P. Li, "Development and Study of $Nb_3Sn$ Wires With High Specific Heat," in IEEE Transactions on Applied Superconductivity, vol. 29, no. 5, Aug. 2019, Art no. 6000404.
16. A. Devred, "Quench Origins," KEK Report 89-25, March 1990.
17. M. K. Thompson and J. M. T., "ANSYS Mechanical APDL for Finite Element Analysis," 1st edition, Butterworth-Heinemann, 2017.
18. X. Xu, A. V. Zlobin, E. Barzi – Fermilab; C. Buehler, M. Field, B. Sailer, M. Wanior, H. Miao – Bruker EST; C. Tarantini – Florida State University. "Enhancing specific heat of $Nb_3Sn$ conductors to improve stability and reduce training". Presented at CEC-ICMC 2019.
19. Bonura, M., and C. Senatore. "Thermal Conductivity of Industrial $Nb_3Sn$ Wires Fabricated by Various Techniques," IEEE Transactions on Applied Superconductivity 23.3 (2013): 6000404. Crossref. Web.
20. G. Manfreda, C. Barbagallo, "Review of ROXIE's Material Properties Database for Quench Simulation," CERN, Magnets, Superconductors and Cryostats TE-MSC Internal Note 2018-1178007.
21. N.J. Simon, E.S. Drexler and R.P. Reed, "Properties of Copper and Copper Alloys at Cryogenic Temperatures," NIST Monography 177, Ed.1992.
22. F. Rondeaux, Ph. Bredy and J.M. Rey, "Thermal Conductivity Measurements of Epoxy Systems at Low Temperatures", I-01C-02, Poster presented at Cryogenic Engineering Conference (CEC), July 16-20, 2001 Madison, Wisconsin, USA.
23. Swenson, C.A., "Linear thermal expansivity (1.5–300 K) and heat capacity (1.2–90 K) of Stycast 2850FT," Review of Scientific Instruments, 1997.

24. Dargis, Rytis et al., "Structural and Thermal Properties of Single Crystalline Epitaxial $Gd_2O_3$ and $Er_2O_3$ Grown on Si(111)." ECS Journal of Solid State Science and Technology. 1. N24. 10.1149/2.005202jss (2012).
25. T. Summers, M. W. Guinan, J. R. Miller, and P. A. Hahn, IEEE Trans. Mag., vol. 27, p. 2041-4, 1991.
26. A. Stenvall, A. Korpela, J. Lehtonen, R. Mikkonen, "Formulation for solving 1D minimum propagation zones in superconductors," Physica C: Superconductivity and its Applications, Volume 468, Issue 13, 2008, Pages 968-973.